\documentclass[twoside,a5paper,11pt]{article}
\usepackage{iwoc-a5}

\title{\Large \bf Anisotropic flow generated by hard partons\\ in medium}
\author{Boris Tom\'a\v{s}ik$^{1,2}$ and Martin Schulc$^2$}
\date{}
\begin{document}
\maketitle

\begin{center}
\vspace*{-0.3cm}
{\it  $^1$~Univerzita Mateja Bela, Tajovsk\'eho 40, 97401 Bansk\'a Bystrica, Slovakia\\
$^2$~Czech Technical University in Prague, FNSPE, B\v{r}ehov\'a 7, \\11519 Praha 1, Czech Republic}\\
e-mail: boris.tomasik@umb.sk
\end{center}

\vspace{0.3cm}

\begin{center}
{\bf Abstract}\\
\medskip
\parbox[t]{10cm}{\footnotesize
Hard partons which are produced copiously in nuclear collisions at the LHC, deposit 
most of their energy and momentum into the surrounding quark-gluon plasma. We show 
that this generates streams in the plasma and contributes importantly to 
flow anisotropies. With the help of event-by-event  
three-dimensional perfect hydrodynamic simulations we 
calculate the observable azimuthal anisotropies of hadronic distributions and show that 
the proposed mechanism is capable of generating non-negligible part of the observed signal. 
Hence, it must be taken into account in quantitative studies in which one tries to extract 
the values of viscosities from the comparison of simulated results with measured data. }
\end{center}


\section{Introduction} 
\label{intro}

Expansion of matter excited in ultrarelativistic nuclear collisions provides access to its 
collective properties: Equation of State and transport coefficients. More detailed study 
of them 
is possible if one looks at azimuthal anisotropies of hadron distributions. They are 
caused by anisotropic expansion of the fireball (for reviews see e.g. 
\cite{roma,gale}). 

Indeed, the slope of a transverse-momentum hadron spectrum is influenced by 
transverse expansion through a Doppler blue-shift. If we select particles  with certain 
momentum there is a specific part of the expanding fireball which dominates the production 
of this momentum. Most naturally this would be that part of the fireball which co-moves 
with the hadrons within our focus. 
Emission of this momentum from other parts---i.e.\ those moving 
in other directions---is suppressed.  Thus we have radiating source moving towards 
the detector. The blue-shift of the radiation is translated into enhanced production 
of higher $p_t$. Therefore the spectrum of an expanding source becomes flatter. 
The range of source velocities co-determines---together with the temperature---the 
flatness of the spectrum.

If the fireball expands with different velocities in different directions this is usually put 
into connection with inhomogeneities in the initial state determined by various kinds
of fluctuations of energy deposition during the initial impact. By hydrodynamically propagating 
these inhomogeneities and comparing thus calculated hadron distributions with measured 
data one tries to determine the properties of matter which enter the evolution model. 

One of the problems with this programme is that the initial conditions are only known 
from various model calculations. Moreover, any other mechanism which influences the 
flow anisotropies hinders the  determination of transport coefficients and must be controlled 
in good quantitative studies. 

We propose here another mechanism which clearly leads to anisotropy in the collective 
expansion of the fireball. It must be well understood quantitatively if further progress 
in determination of matter properties is desired. 


\section{Flow anisotropy from hard partons}
\label{s:flowaniso}

In nuclear collisions at collider energies non-negligible part of the energy is initially 
released in the form of partons with high transverse momentum. At the LHC we have several pairs 
of partons with transverse momentum above a few GeV per event. Usually, one would refer 
to them as seeds of minijets and jets. In most cases all of their momentum and energy, 
however, is transferred into the surrounding quark-gluon-plasma over some period of time. 
This creates streams in the bulk which lead to anisotropies in  collective expansion. 

It is quite conceivable that such a mechanism can lead to flow ani\-so\-tro\-pies 
which fluctuate from 
event to event. Below we will estimate the effect with the help of hydrodynamic simulation. 
\textit{A priori}
it is not clear, however, whether this mechanism is oriented fully randomly or whether 
it is correlated with the geometry of the collision and also contributes to event-averaged
anisotropies. 

\begin{figure}[htb]
\begin{center}
\includegraphics[width=0.3\textwidth]{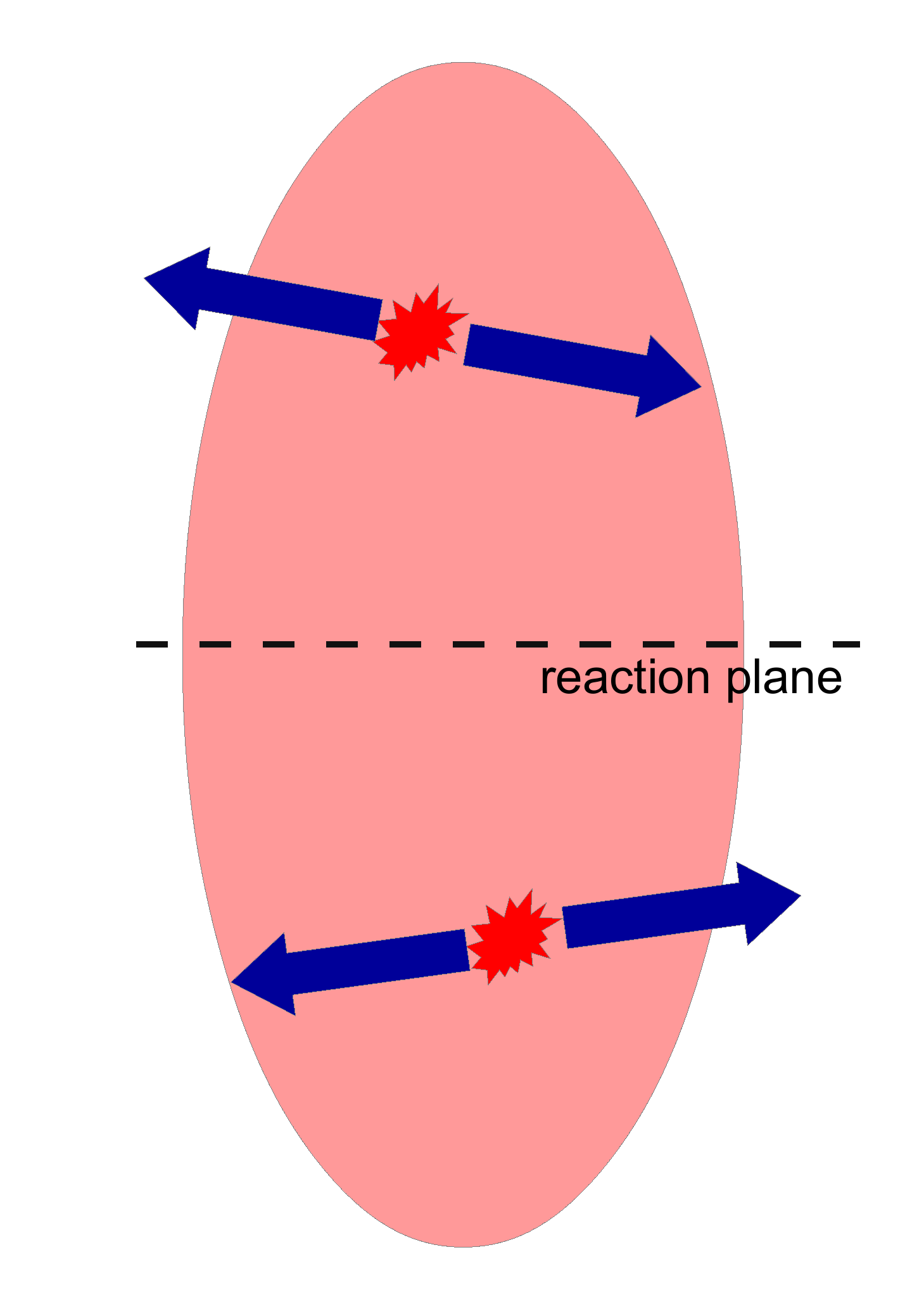}
\hspace{1cm}
\includegraphics[width=0.3\textwidth]{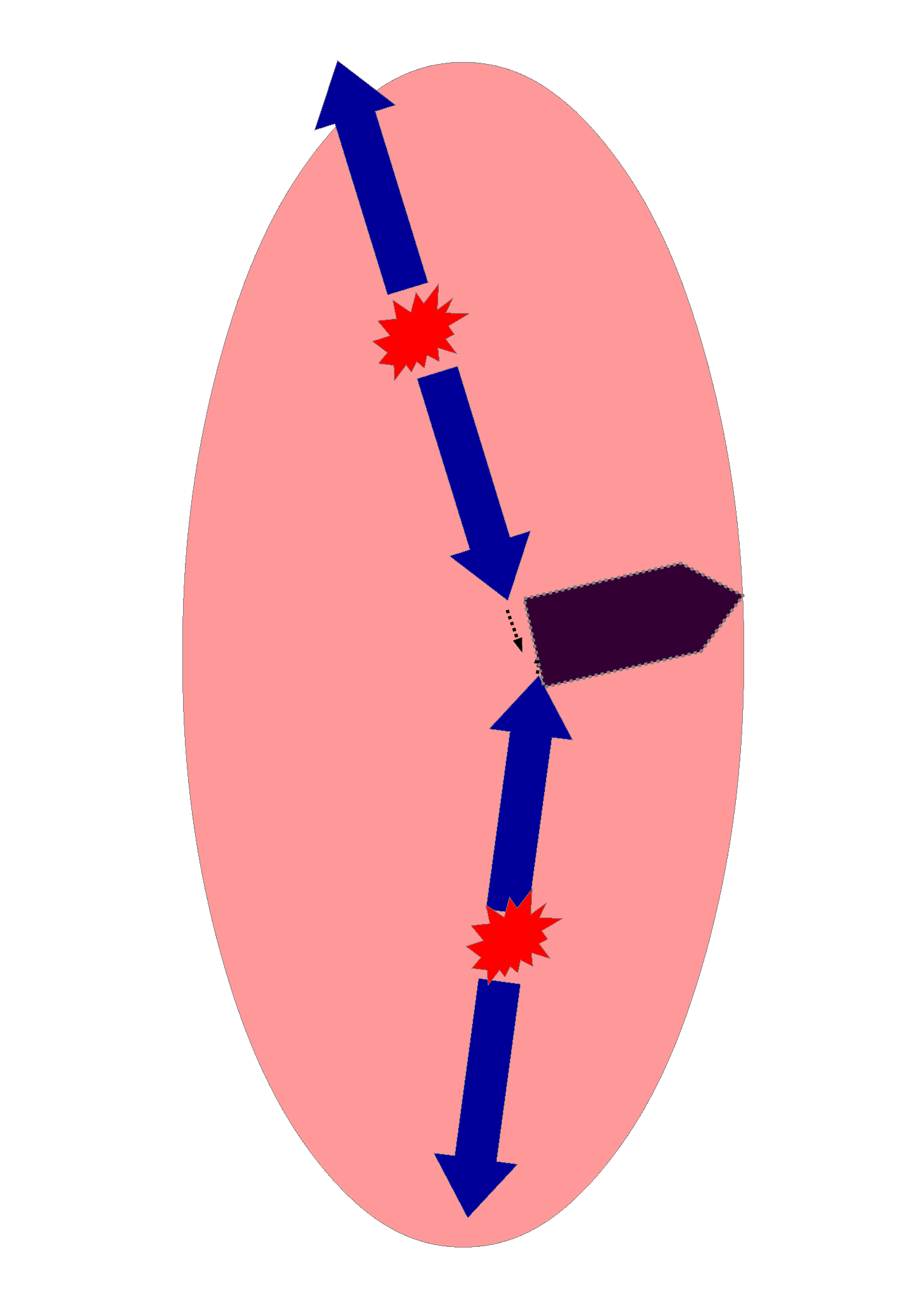}
\end{center}
\vspace{-0.5cm} 
\caption{Left: two dijets produced both almost parallel to the reaction plane. 
Blue arrows represent generated streams within 
quark-gluon plasma. Right: Two dijets  produced in directions out of reaction plane. } 
\label{f:cartoon}
\end{figure}

The latter is the case. Let us first explain how the mechanism works. In a non-central 
collision the fireball is initially elongated in the direction perpendicular to the reaction 
plane (which is spanned by the beam direction and the impact parameter). If two 
dijets are produced and directed both along the reaction plane, they both contribute 
to the elliptic flow anisotropy, as pictured in Fig.~\ref{f:cartoon} left. 
Their contribution is positive, because due 
to larger pressure gradient stronger flow usually develops along the reaction plane
even without any hard partons. 
If, however, two hard partons are directed out of the reaction plane, the chance is higher 
that two jet-induced streams will meet. Then they merge and continue in direction determined
by energy and momentum conservation, see Fig.~\ref{f:cartoon}. Such a merger is more likely 
in this case since here the streams pass each other along the narrow 
direction of the fireball and have less space  to avoid the merger. 

Note that via this mechanism isotropically produced hard partons couple to anisotropic 
shape of the fireball and generate anisotropy of the collective expansion. 

An early study mimicking such a mechanism indicated that it will lead to elliptic flow, indeed
\cite{Tomasik:2011xn}. 


\section{Hydrodynamic simulations} 
\label{s:hydro}

In order to test our ideas in more realistic simulations we have constructed 
3D hydrodynamic model \cite{schulc}. We assumed perfect fluid. Simu\-lations
including viscosity are planned for the future. Note that it is important that the
simulation is three-dimensional. Lower-dimensional models assume some kind of 
symmetry: boost-invariance in case of 2D and additional azimuthal symmetry in case 
of 1D. Inclusion of hard partons, however, breaks 
these symmetries and thus full simulation is needed. 

Hard partons may deposit large amount of energy into a small volume and its evolution may lead 
to shock waves. Thus the model must exploit an algorithm capable of handling such a situation. 
We use SHASTA \cite{Boris,DeVore}. 

First we have shown that in a static medium hard partons induce streams which can merge 
if they come into contact \cite{schulc}. 

Then we ran simulations of collisions of Pb nuclei at full LHC energy 
$\sqrt{s_{NN}} = 5.5$~TeV. Initial energy density profile is smooth and follows from 
the optical Glauber model. The initial positions of hard partons follow the distribution
of binary  nucelon-nucleon collisions. Energy and momentum deposition from hard partons  into 
plasma is described as $\partial_\mu T^{\mu\nu} = J^\nu$ with the force term 
\begin{equation}
J^\nu = -\sum_i \frac{1}{(2\, \pi\, \sigma_i^2)^{\frac{3}{2}}} \, \exp \left (
- \frac{\left ( \vec x - \vec x_{\mathrm{jet},i} \right )^2 }{2\, \sigma_i^2} \right )\, 
\left ( \frac{dE_i}{dt}, \frac{d\vec P_i}{dt} \right )
\end{equation}
where the sum runs over all hard partons in the event. We did not study the microscopic 
mechanism  of energy transfer from hard partons to plasma and only assumed 
that it is localised within Gaussian distribution with $\sigma_i = 0.3$~fm. 

The energy loss per unit of length scales with the entropy density 
\begin{equation}
\frac{dE}{dt} \approx c\frac{dE}{dx}\, , \qquad \frac{dE}{dx} = \frac{s}{s_0}\frac{dE}{dx}\Biggr |_0
\end{equation}
where $s_0$ corresponds to energy density of 20~GeV/fm$^3$ and $dE/dx|_0$ 
is a parameter of the simulation for which we tested a few values. Details of the model
can be found in \cite{svieze_dielko}.


\section{Results}
\label{s:results}

Due to flow fluctuations  flow anisotropies are generated even in most central 
collisions. They are observable if one does not average over many events. We first looked 
at the contribution of our mechanism to anisotropies in central collisions. To this end 
we simulated 100 central events with included hard partons and then ran THERMINATOR2
\cite{THERM2} five times in order to generate hadrons  
for each of the obtained freeze-out hypersurfaces. 

\begin{figure}[htb!]
\begin{center}
\includegraphics[width=0.65\textwidth]{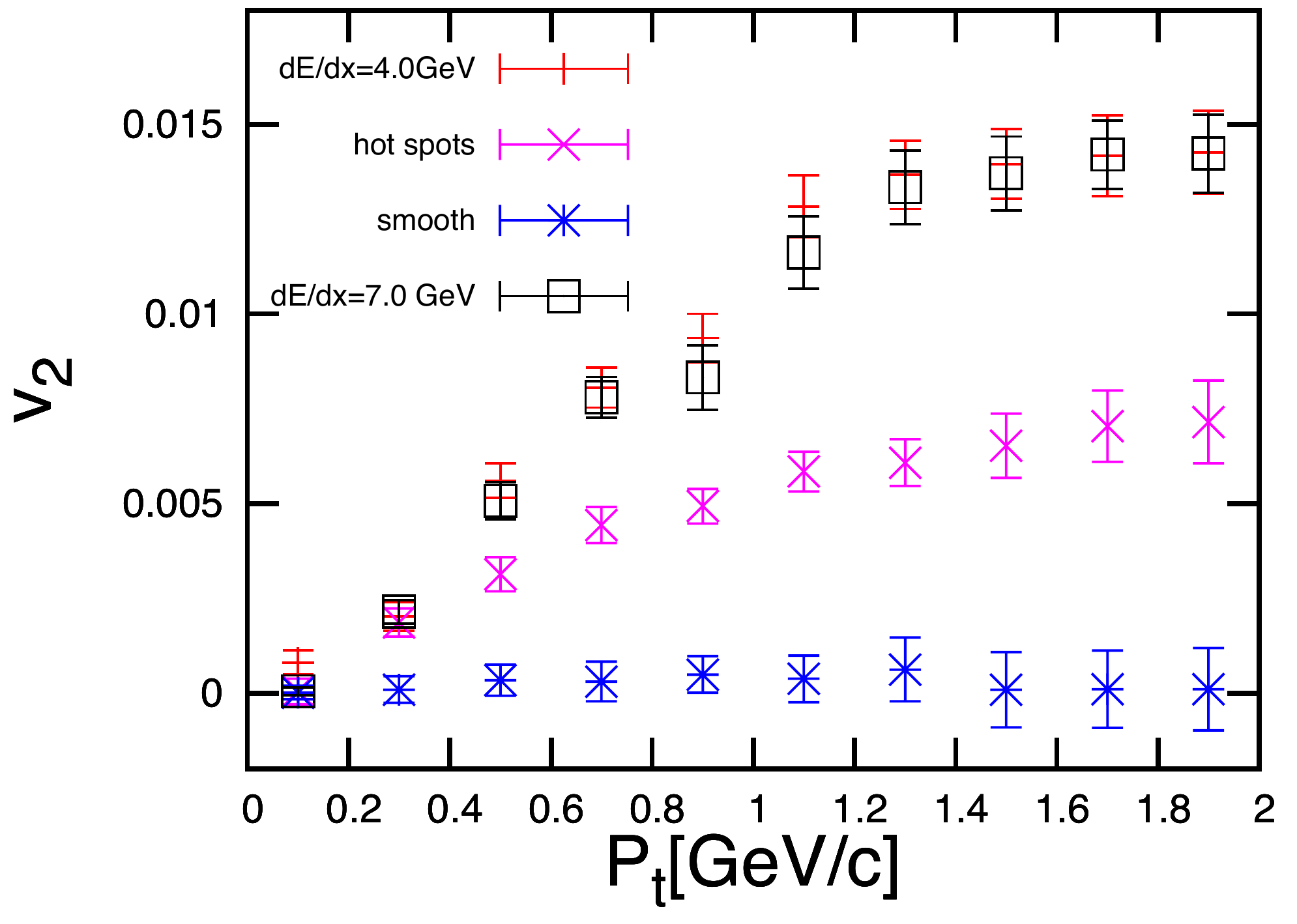}
\includegraphics[width=0.65\textwidth]{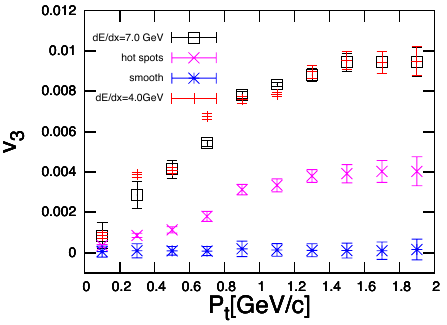}
\end{center}
\vspace{-0.5cm} 
\caption{Top: Second order anisotropy coefficient $v_2$ of hadronic distributions in 
ultra-central collisions. Lower data are from simulation with no fluctuations. 
Upper two sets of data are from simulations with hard partons with different 
values of $dE/dx|_0$. Crosses in between of these data sets show results 
from simulations with hot spots instead of hard partons. Bottom: same as 
top panel but for $v_3$.
} 
\label{f:central}
\end{figure}

In Fig.~\ref{f:central} we show 2nd and 3rd order anisotropy coefficients $v_2$ and 
$v_3$. 
Results are compared to simulations with no hard partons which indeed show 
no anisotropies. We studied the dependence on the value of $dE/dx|_0$. 
Surprisingly results seem not to depend on the particular value of the energy loss.
Note that the total amount of the energy deposited into plasma is the same in both cases. 
They differ by how fast this process runs. The reason may be that in most cases 
all energy is deposited from hard partons into plasma already at the beginning. 
We also measure the anisotropies in simulations where hard partons were replaced 
by hot spots, i.e.\ local depositions of additional energy density in the initial conditions. 
They are chosen in such a way that onto the smooth energy density 
profile  the same amount of energy is added as the hard partons would deposit during 
the whole time. We see that the effect generating flow anisotropies is smaller than 
with hard partons where also momentum is deposited. 

As a cross-check, we confirmed that no anisotropies of hadron spectra are generated from 
azimuthally symmetric fireball with smooth initial conditions and no fluctuations. 

\begin{figure}[htb]
\begin{center}
\includegraphics[width=0.65\textwidth]{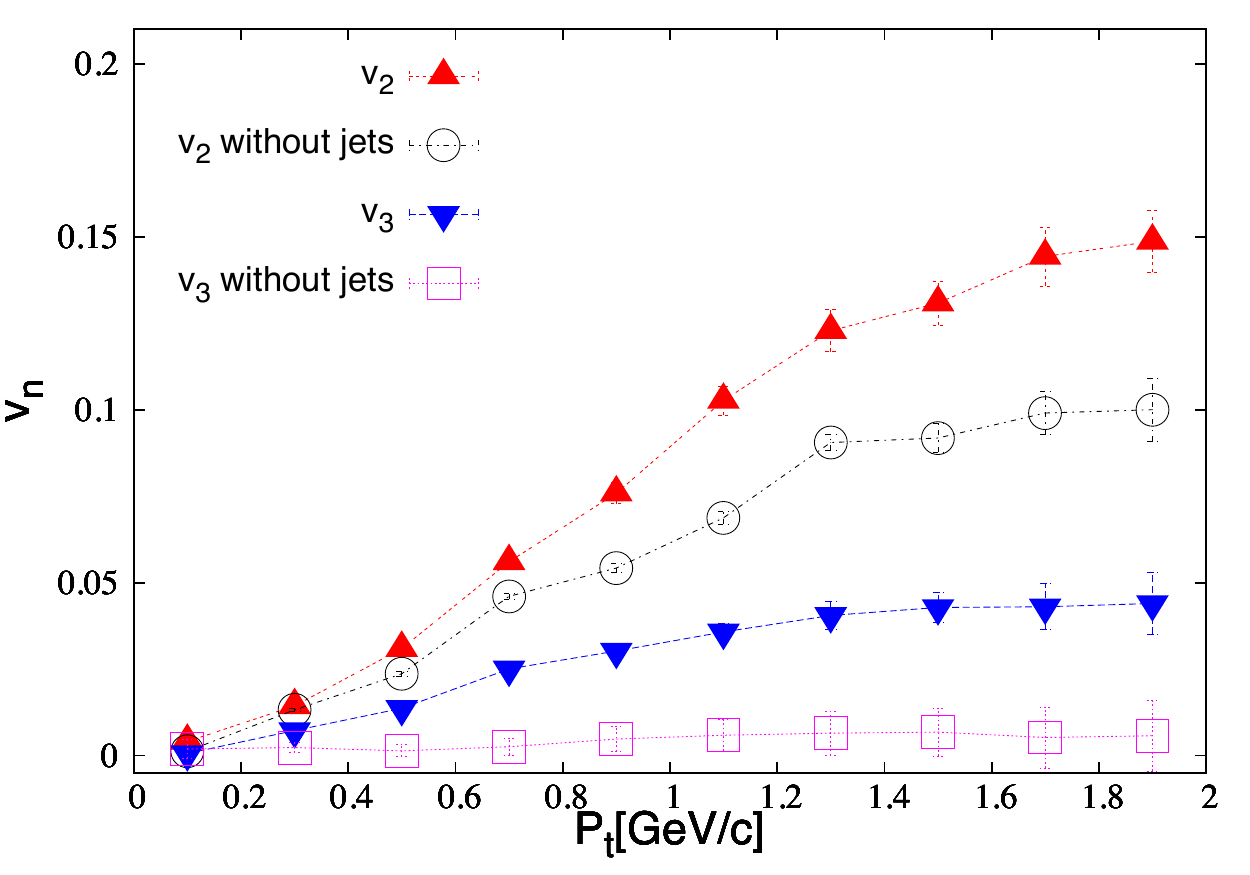}
\end{center}
\vspace{-0.5cm} 
\caption{%
Azimuthal anisotropy coefficients $v_2$ and $v_3$ from simulations of 
30-40\% centrality class (impact parameter 
$b = 6.5$~fm). Simulations with hard partons  
are compared to simulations with only smooth initial conditions and no hard 
partons (without jets). 
} 
\label{f:noncen}
\end{figure}

We also checked how this mechanism is  aligned with the geometry in non-central 
collisions. To this end we simulated fireballs with impact parameter $b=6.5$~fm
and compared anisotropies in cases with or without hard partons, see Fig.~\ref{f:noncen}.
Hard partons indeed enhance the elliptic flow; this confirms the alignment with collision 
geometry thanks to merging of the streams. Triangular flow ($v_3$) is solely generated by 
hard partons. It is absent in non-central collisions with smooth initial conditions
in accord with the symmetry constraints.


\section{Conclusions}
\label{s:conc}

There are several studies similar to ours documented in the literature. 

In \cite{Tachibana} the authors study the response of expanding fireball to only 
one dijet. As we argued previously, this cannot lead to the alignment with the geometry 
since it is caused by merging of the induced streams. 

Simulations in \cite{Noronha} are performed in 2D. We argued that using boost-invariance
in this case may be inappropriate. 

Finally, in \cite{Florchinger,Zapp} the authors only study the influence of hard partons 
on radial flow and did not touch elliptic flow anisotropies.

Our results show that the contribution to flow anisotropies from hard partons
may be relevant in quantitative studies aimed at the determination of the transport 
coefficients. More precise studies will require inclusion of three-dimensional 
viscous hydrodynamic model and other sources of fluctuations. 


\paragraph{Acknowledgement}
This work was supported in parts by APVV-0050-11, VEGA 1/0457/12 (Slovakia) and 
M\v{S}MT grant  LG13031 (Czech Republic).



\begin{thebibliography}{99}

\bibitem{roma}
P.\ Romatschke, Int.\ J.\ Mod.\ Phys.\ E \textbf{19} 1 (2010).

\bibitem{gale}
C.\ Gale, S.\ Jeon, B.\ Schenke, Int.\ J.\ Mod.\ Phys.\ A \textbf{28} 1340011 (2013).

\bibitem{Tomasik:2011xn}
  B.~Tom\'a\v{s}ik and P.~L\'evai,
  J.\ Phys.\ G {\bf 38} 095101 (2011).

\bibitem{schulc}
M.\ Schulc, B.\ Tom\'a\v{s}ik,
 J.\ Phys.\ G {\bf 40} 125104 (2013).

\bibitem{Boris}
  J.~P.~Boris, D.~L.~Book,
  J.~Comp.~Phys.~{\bf 11} 38 (1973).
 
 \bibitem{DeVore}
 C.~R.~DeVore, J.~Comput.~Phys.~{\bf 92} 142 (1991).


\bibitem{svieze_dielko}
M.\ Schulc, B.\ Tom\'a\v{s}ik, Phys. Rev. C \textbf{90} 064910 (2014).

\bibitem{THERM2}
  M.~Chojnacki, A.~Kisiel, W.~Florkowski and W.~Broniowski,
  Comput.\ Phys.\ Commun.\  {\bf 183} 746 (2012).

\bibitem{Tachibana}
Y. Tachibana, T. Hirano, Phys.\ Rev. C \textbf{90} 021902 (2014) .

\bibitem{Noronha}
 R.~P.~G.~Andrade, J.~Noronha and G.~S.~Denicol,
Phys.\ Rev.\ C \textbf{90} 024914 (2014).

\bibitem{Florchinger}
S.~Floerchinger and K.~C.~Zapp,
Eur. Phys. J. C \textbf{74} 3189 (2014).

\bibitem{Zapp}
K.~C.~Zapp and S.~Floerchinger,
  Nucl.\ Phys.\ A\textbf{931} 388 (2014).




\end{thebibliography}
\end{document}